\documentclass{article}

\usepackage[preprint, nonatbib]{neurips_2023}
\usepackage[numbers]{natbib}

\usepackage[utf8]{inputenc} 
\usepackage[T1]{fontenc}    
\usepackage{hyperref}       
\usepackage{url}            
\usepackage{booktabs}       
\usepackage{amsfonts}       
\usepackage{nicefrac}       
\usepackage{microtype}      
\usepackage{xcolor}         
\usepackage{graphicx}

\title{ADMET property prediction through combinations of molecular fingerprints}

\author{%
  James H. Notwell \\
  MapLight Therapeutics\\
  \texttt{jnotwell@maplightrx.com} \\
  \And
  Michael W. Wood \\
  MapLight Therapeutics \\
  \texttt{mwood@maplightrx.com} \\
}

\begin{document}

\maketitle

\begin{abstract}
While investigating methods to predict small molecule potencies, we found random
forests or support vector machines paired with extended-connectivity
fingerprints (ECFP) consistently outperformed recently developed methods. A
detailed investigation into regression algorithms and molecular fingerprints
revealed gradient-boosted decision trees, particularly CatBoost, in conjunction
with a combination of ECFP, Avalon, and ErG fingerprints, as well as 200
molecular properties, to be most effective. Incorporating a graph neural network
fingerprint further enhanced performance. We successfully validated our model
across 22 Therapeutics Data Commons ADMET benchmarks. Our findings underscore
the significance of richer molecular representations for accurate property
prediction.
\end{abstract}

\section{Summary}

We set out to find a potency predictor for small molecules that inhibit a novel
 class A GPCR (subsequently referred to as GPRX) where no crystal structure was
 available. Excited by the developments in applying different neural network
 architectures to molecular property prediction, we applied deep message passing
 neural networks (Chemprop \cite{heid_chemprop_2023}), text-based transformers
 (ChemBERTa \cite{chithrananda_chemberta_2020} and ChemBERTa2
 \cite{ahmad_chemberta-2_2022}), and graph neural networks (Grover
 \cite{rong_self-supervised_2020}) to more than 300 molecules we had synthesized
 for this program. Unfortunately, even models that leveraged pretraining on
 large molecular datasets did not outperform random forests or support vector
 machines coupled with extended-connectivity fingerprints (ECFP)
 \cite{rogers_extended-connectivity_2010}.

This observation led us to investigate different regression algorithms coupled
with different molecular fingerprints. We learned several things from this
exercise: 1. Gradient-boosted decision trees (GBDT) performed better than other
algorithms, and CatBoost \cite{prokhorenkova_catboost_2018} performed slightly
but consistently better than other GBDT implementations,  2. ECFP and Avalon
fingerprints \cite{gedeck_qsar_2006} performed better than other molecular
fingerprints available through RDKit \cite{landrum_rdkitrdkit_2023}, and
modifying the parameters of these fingerprints had little effect on potency
prediction error, and 3. Combining the ECFP and Avalon fingerprints with 200
molecular properties, e.g. number of rings, molecular weight, etc., performed
better than the ECFP and Avalon fingerprints on their own. 

Having now syntesized and assayed hundreds of additional inhibitors of GPRX, we
wanted to see if we could get more performance out of our best performing model
through hyperparameter optimization. To ensure the generalization of our model,
we used a time-split validation \cite{sheridan_time-split_2013} and set aside all
molecules profiled after a given date, as well as those being synthesized, to
serve as a test set. With the remaining molecules, we performed a scaffold
split, reserving 20\% of the molecules for model validation. While
CatBoost was the most accurate GBDT regressor in our testing, we found LigtGBM
\cite{ke_lightgbm_2017} easier to work with when combined with the Optuna
\cite{akiba_optuna_2019} hyperparameter optimization framework, including having
its own tuner \cite{ozaki_lightgbm_2022}. We explored a number of
cross-validation strategies, including grouping molecules with similar scaffolds
or potencies into the same or different folds, as well as randomizing across
these properties, but observed consistent overfitting across extensive
hyperparameter searching. When evaluating the validation set, we found it
difficult to outperform LightGBM or CatBoost with default parameters, with the
sole exception of modifying the random strength in CatBoost.

The idea of combining molecular features is not new. Concatenating featurizers
is supported in the molfeat package \cite{noutahi_datamol-iomolfeat_2023} and
was used in the top submission \cite{kengkanna_enhancing_2023} for the 1st
EUOS/SLAS Joint Challenge to predict compound solubility
\cite{andrea_zaliani_1st_2022}. Intrigued by our initial results, we leveraged
molfeat featurizers to explore all combinations of 1, 2, or 3 featurizers. This
search highlighted a third type of fingerprint that performed well when used in
conjunction with ECFP and Avalon fingerprints: extended reduced graph approach
(ErG) \cite{stiefl_erg_2006}. Interestingly, the neural network fingerprints,
regardless of architecture type, performed worse compared to traditional
molecular fingerprints. Having performed extensive searches across model space,
we applied our model to the test portion of our time split and observed
excellent generalization with error similar to the validation set. We wondered
if our model's performance would extend to other molecular property prediction
tasks. 

\begin{figure}
  \centering
  \includegraphics[width=\linewidth]{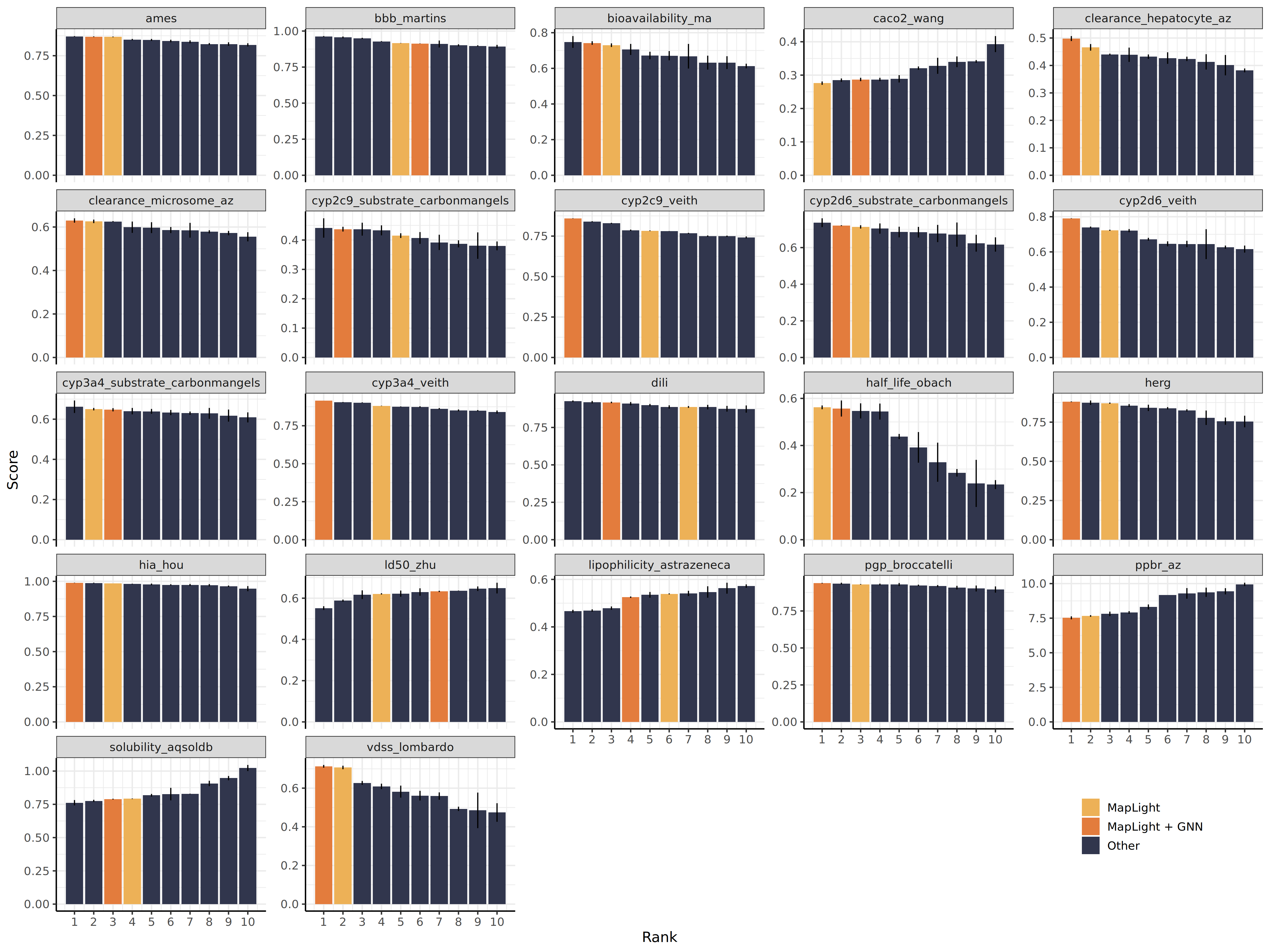}
  \caption{Relative performance of the MapLight model, with (orange) and without (yellow) GNN fingerprints, compared to the current TDC leaderboards (dark blue). Bar heights and error bars correspond to the mean and standard deviation across 5 random seeds.}
  \label{fig:Fig1}
\end{figure}

The Therapeutics Data Commons (TDC) has compiled 22 benchmarks across different
ADMET property prediction tasks, such as aqueous solubility or cytochrome P450
enzyme inhibition \cite{huang_artificial_2022}. To further probe its
generalization, we applied our final model, consisting of a CatBoost classifier
or regressor coupled with molecular representations that were a combination of
ECFP, Avalon, and ErG fingerprints, as well as 200 molecular properties, to
these benchmarks. When doing so, our model achieved top-1 performance in 6 of 22
benchmarks and top-3 performance in 16 of 22 benchmarks (Figure \ref{fig:Fig1}).
To test the limits of combining fingerprints, we added a graph isomorphism
network (GIN - supervised masking variant \cite{hu_strategies_2020}) fingerprint
to our molecular representation, and achieved top-1 performance in 11 of 22
benchmarks and top-3 performance in 19 of 22 benchmarks (Figure \ref{fig:Fig1}).

Lurking in the background of longer and longer molecular descriptors is the
curse of dimensionality. Given the generalization of a single model across 22
benchmarks, as well as to the time-split of our internally generated GPRX
inhibitors, the performance of our model is unlikely the result of overfitting.
While we combine fingerprints, our total molecular representation is less than
50\% larger than common implementations of ECFP fingerprints (ECFP counts
(length 1024) + Avalon counts (length 1024) + ErG (length 315) + molecular
properties (length 200) + GIN supervised masking (length 300) = length 2863). We
modify a single CatBoost hyperparameter (random strength), although
qualitatively similar performance can be achieved with other GBDT
implementations, e.g. LightGBM, with default parameters. In comparison with
other approaches, such as those from \citet{huang_unified_2022}, we achieve
strong results with a single model.

A recent editorial was titled "For Chemists, the AI revolution has yet to
happen" \cite{noauthor_for_2023}. While this article places much of the blame on
the availability of training data, a contributing factor is how we describe
molecules. A key finding from our work is that richer molecular representations,
achieved through combining fingerprints, provide strong predictive power.
Looking across the 22 TDC ADMET leaderboards, we see additional evidence for
this: the deep message passing neural network (Chemprop) is a generally good
performer but does even better when combined with molecular properties
(Chemprop-RDKit). This is a promising avenue for future work.

\section{Software Availability}

The software for combining molecular fingerprints can be found at
\url{https://github.com/maplightrx/MapLight-TDC} and is released under the MIT
License.

\bibliographystyle{abbrvnat}
\bibliography{references}

\end{document}